\documentclass[aps, prx, reprint, superscriptaddress]{revtex4-2}

\pdfoutput=1 
\usepackage{graphicx}
\usepackage{color,amsmath}
\usepackage[normalem]{ulem}
\usepackage[utf8]{inputenc}

\usepackage[colorlinks,linkcolor=blue,citecolor=blue,urlcolor=blue]{hyperref}
\usepackage[separate-uncertainty = true]{siunitx}
\usepackage{comment}
\usepackage{layouts}
\usepackage[acronym]{glossaries}
\glsdisablehyper

\begin{document}
\title{Fractional Coulomb blockade for quasiparticle tunneling between edge channels}

\author{Marc~P.~Röösli}
\email{marcro@phys.ethz.ch}
\affiliation{Solid State Physics Laboratory, Department of Physics, ETH Zurich, 8093 Zurich, Switzerland}

\author{Michael~Hug}
\affiliation{Solid State Physics Laboratory, Department of Physics, ETH Zurich, 8093 Zurich, Switzerland}

\author{Giorgio~Nicolí}
\affiliation{Solid State Physics Laboratory, Department of Physics, ETH Zurich, 8093 Zurich, Switzerland}

\author{Peter~Märki}
\affiliation{Solid State Physics Laboratory, Department of Physics, ETH Zurich, 8093 Zurich, Switzerland}

\author{Christian~Reichl}
\affiliation{Solid State Physics Laboratory, Department of Physics, ETH Zurich, 8093 Zurich, Switzerland}

\author{Bernd~Rosenow}
\affiliation{Institute for Theoretical Physics, Leipzig University, Leipzig D-04009, Germany}

\author{Werner~Wegscheider}
\affiliation{Solid State Physics Laboratory, Department of Physics, ETH Zurich, 8093 Zurich, Switzerland}

\author{Klaus~Ensslin}
\affiliation{Solid State Physics Laboratory, Department of Physics, ETH Zurich, 8093 Zurich, Switzerland}

\author{Thomas~Ihn}
\affiliation{Solid State Physics Laboratory, Department of Physics, ETH Zurich, 8093 Zurich, Switzerland}

\date{\today}

\newacronym[longplural={quantum dots}]{QD}{QD}{quantum dot}
\newacronym[longplural={quantum point contacts}]{QPC}{QPC}{quantum point contact}
\newacronym[longplural={charge stability diagrams}]{CSD}{CSD}{charge stability diagram}
\newacronym[longplural={two-dimensional electron gases}]{2DEG}{2DEG}{two dimensional electron gas}

\begin{abstract}
We study the magneto-conductance of a \SI{1.4}{\micro m}-wide quantum dot in the fractional quantum Hall regime. For a filling factor $\approx 2/3$  and $\gtrsim 1/3$ in the quantum dot the observed Coulomb resonances show a periodic modulation in magnetic field. This indicates a non-trivial reconstruction of the 2/3 fractional quantum Hall state in the quantum dot.  We present a model for the charge stability diagram of the system assuming two compressible regions separated by an incompressible stripe of filling factor $2/3$ and $1/3$, respectively. From the dependence of the magnetic field period on total magnetic field we construct the zero-field charge density distribution in the quantum dot. The tunneling between the two compressible regions exhibits fractional Coulomb blockade. For both filling factor regions, we extract a fractional charge $e^*/e = \SI{0.32 \pm 0.03}{}$ by comparing to measurements at filling factor 2. With their close relation to quantum Hall Fabry-P\'{e}rot interferometers, our investigations on quantum dots in the fractional quantum Hall regime extend and complement interference experiments investigating the nature of anyonic fractional quantum Hall quasiparticles.
\end{abstract}

\maketitle
Large \glspl{QD} can be used to study physical processes in the quantum Hall regime. Nevertheless no effects of quasiparticle tunneling have been observed for Coulomb blockaded \glspl{QD} in the fractional quantum Hall regime so far \cite{sabo_edge_2017}. The reason is that tunneling barriers connecting a \gls{QD} in the Coulomb blockade to source and drain regions are strongly backscattering and therefore only allow for electron tunneling \cite{chang_chiral_2003}. Operating barriers in the regime of weak backscattering allows for quasiparticle tunneling. Evidence for fractionally charged quasiparticles has previously been observed in experiments on shot noise of a \gls{QPC} \cite{de-picciotto_direct_1997,saminadayar_observation_1997,reznikov_observation_1999,dolev_observation_2008,bid_shot_2009}, capacitively probed localized states \cite{martin_localization_2004}, anti-dots\cite{kou_coulomb_2012} or quantum Hall Fabry-P\'{e}rot interferometers \cite{nakamura_aharonovbohm_2019}. In general, quantum Hall Fabry-P\'{e}rot interferometer experiments \cite{van_wees_observation_1989,taylor_aharonov-bohm_1992,bird_precise_1996,godfrey_aharonov-bohm-like_2007,camino_aharonov-bohm_2005-1} in the fractional quantum Hall regime offer the opportunity to study anyonic statistics of fractional quasiparticles \cite{ofek_role_2010,camino_aharonov-bohm_2005,zhou_flux-period_2006,camino_$e/3$_2007,camino_quantum_2007,lin_electron_2009,mcclure_fabry-perot_2012,willett_measurement_2009,willett_alternation_2010,nakamura_aharonovbohm_2019,willett_interference_2019}. The close relation between \glspl{QD} and Fabry-P\'{e}rot interferometers in the quantum Hall regime \cite{rosenow_influence_2007,stern_interference_2010,halperin_theory_2011,sivan_observation_2016,roosli_observation_2020} promises complementary experimental observations on fractional quasiparticles in \glspl{QD}. 

In this paper, we study the magneto-transport through a large \gls{QD} containing roughly 900 electrons in the fractional quantum Hall regime for filling factors $\nu < 1$. The \gls{QD} forms concentric compressible regions separated by incompressible regions. In the integer quantum Hall regime this has been established by experiments \cite{brown_resonant_1989,mceuen_transport_1991,mceuen_self-consistent_1992,mceuen_coulomb_1993} and theory \cite{dempsey_electron-electron_1993}. By reconstructing the charge carrier distribution in the \gls{QD} at zero magnetic field we can show that in two specific regions of magnetic field, the incompressible region corresponds to a fractional filling factor $\nu_\mathrm{in} = 1/3$ or $2/3$, respectively. In our experiments, the \gls{QD} is weakly tunnel-coupled to its leads and occupied by an integer number of electrons $N$. In this regime conductance peaks arise each time the chemical potential of the $N$-th electron state and the leads are degenerate, as usual in Coulomb blockade experiments. While only an integer number of electrons can tunnel between the \gls{QD} and the leads \cite{chang_chiral_2003}, we observe fractional quasiparticle tunneling between the compressible regions inside the \gls{QD} with a fractional charge corresponding to $e^* = e/3$ as predicted theoretically. In the properly tuned regime the tunnel-coupling across the incompressible region is strong enough to enable tunneling of fractionally charged quasiparticles, and weak enough to lead to a detectable Coulomb blockade signal via rearrangement of (fractional) charges. As each compressible region forms a \gls{QD}, the quasiparticle tunneling can be treated in a capacitive single particle model as the fractional Coulomb blockade between two nested \glspl{QD}. We construct a phase diagram of stable charge which was previously proposed theoretically \cite{evans_coulomb_1993} and measured for integer Landau levels \cite{,van_der_vaart_time-resolved_1994,van_der_vaart_time-resolved_1997,heinzel_periodic_1994,fuhrer_transport_2001,chen_transport_2009,liu_electrochemical_2018,roosli_observation_2020}.

\begin{figure}[htbp]
\includegraphics[width=\columnwidth]{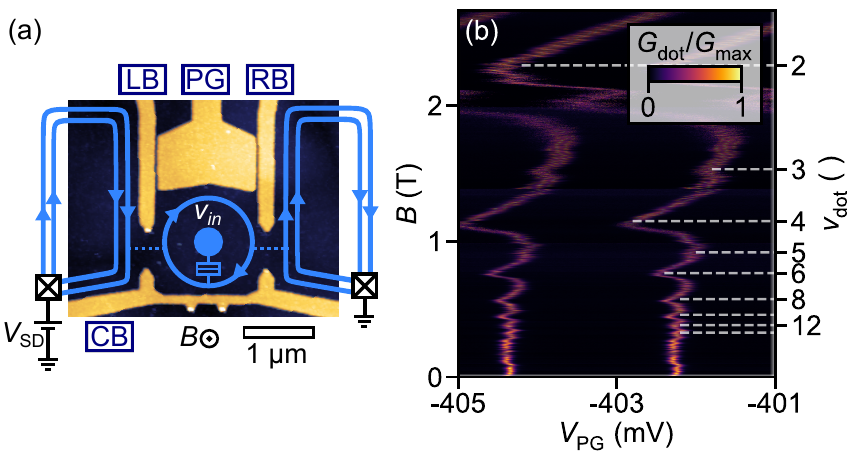}
\caption{(a) Atomic force microscope (AFM) image  of the \acrfull{QD} device. Top gates appear in yellow while the uncovered semiconductor is dark blue. A magnetic field is applied perpendicular to the sample surface. The overlaid schematics show the chiral compressible regions for a bulk filling factor $\nu_\mathrm{b} \approx 2$. We apply  a source-drain bias $V_\mathrm{SD}$ and measure the current $I_\mathrm{SD}$. (b) Normalized conductance $G_\mathrm{dot}/G_\mathrm{max}$ through the \gls{QD} as a function of the plunger gate voltage $V_\mathrm{PG}$ and the magnetic field $B$, where $G_\mathrm{max}$ is the local maximum of the conductance over 5 adjacent plunger gate voltage traces. The Coulomb peaks show a $1/B$-periodic behavior that can be related to integer filling factors $\nu_\mathrm{dot} = 2,\,3,\,4,\,5,\,6,\,8,\,10,\,12$ and 14 in the \gls{QD} (dashed lines). Fitting these features results in an electron density $n_\mathrm{dot} = \SI{1.11 \pm 0.04e11}{cm^{-2}}$ in the \gls{QD}. A corresponding filling factor ($\nu_\mathrm{dot}$) axis is indicated on the right.} 
\label{fig1}
\end{figure}

The \gls{QD} sample is fabricated on an AlGaAs/GaAs-heterostructure etched into a Hall bar structure. It hosts a \gls{2DEG} \SI{130}{nm} below the surface which is contacted by annealed AuGeNi ohmic contacts. We measure a bulk electron density $n_\mathrm{bulk} = \SI{1.44e11}{cm^{-2}}$ and electron mobility $\mu = \SI{5.6e6}{cm^{2}/Vs}$ at temperature $T=\SI{30}{mK}$. The bulk electron density can be altered by applying a voltage to the pre-patterned, overgrown back gate extending underneath the Hall bar \SI{1}{\micro m} below the \gls{2DEG} \cite{Berl_Structured_2016}. For all measurements in this paper the back gate was grounded. However, the presence of back gates or additional gates can influence the confinement potential which was exploited in previous experiments \cite{nakamura_aharonovbohm_2019,willett_interference_2019,choi_robust_2015,sivan_interaction-induced_2018}.

The detailed gate design of the inner structure of the \gls{QD} sample is shown in Fig.~\ref{fig1}(a). The \gls{QD} with a width of $\SI{1.4}{\micro m}$ is formed by four metallic gates (labelled CB, LB, RB, and PG) that are lithographically patterned  on the surface of the AlGaAs/GaAs-heterostructure (dark blue). We form the \gls{QD} by applying negative voltages to the gates thereby depleting the electron gas underneath. Depletion of the electron gas below the gate occurs at \SI{-0.35}{V}. By applying a small voltage $V_\mathrm{SD}$ between the source and drain contacts and measuring the resulting current $I_\mathrm{SD}$, we study the two-terminal linear conductance $G_\mathrm{dot} = V_\mathrm{SD}/I_\mathrm{SD} $ of the \gls{QD}. First, we tune the \gls{QD} system into the Coulomb blockade regime. The center barrier (CB) gate and the left and right barrier (LB and RB) gate tune the transmission of the right an left barrier respectively. The transmission of both barriers is set to $\ll e^2/h$ such that the \gls{QD} is only weakly coupled to its leads. We fix the voltage of the CB gate $V_\mathrm{CB} = \SI{-1.2}{V}$ while the voltage on LB and RB is changed to retune the barrier coupling of the \gls{QD} for different measurements. The voltage on the plunger gate (PG) is varied around $V_\mathrm{CB} \approx \SI{-0.4}{V}$ and used to tune the discrete energy levels of the \gls{QD}.

The measurements were conducted in a dilution refrigerator at the base temperature $T = \SI{30}{mK}$. All measurements presented within this paper were performed on the same sample during one cooldown. However, similar measurements were reproduced with another sample using a different heterostructure and gate design.

The electron density in the \gls{QD} region is reduced compared to the bulk density of the \gls{2DEG} by the applied confining gate voltages. In order to estimate the effective electron density $n_\mathrm{dot}$ inside the \gls{QD} we analyze the conductance $G_\mathrm{dot}$ through the \gls{QD}  in the integer quantum Hall regime. Fig.~\ref{fig1}(b) shows the normalized conductance $G_\mathrm{dot}/G_\mathrm{max}$ as a function of magnetic field $B$ applied perpendicular to the sample surface and the plunger gate voltage $V_\mathrm{PG}$. We observe Coulomb blockade resonances as a function of the PG voltage. Their position in gate voltage shows a $1/B$-periodic modulation as indicated by dashed lines in Fig.~\ref{fig1}(b). Identifying these features with integer filling factors $\nu_\mathrm{dot} = 2,\,3,\,4,\,5,\,6,\,8,\,10,\,12$ and $14$ in the \gls{QD} allows us to extract the electron density $n_\mathrm{dot} = \SI{1.11 \pm 0.04e11}{cm^{-2}}$ in the \gls{QD} by fitting the relation $\nu = n_\mathrm{dot} h/(eB)$. We will see later that this density corresponds to the maximum local density in the quantum dot center.

\begin{figure}[tbp]
\includegraphics[width=\columnwidth]{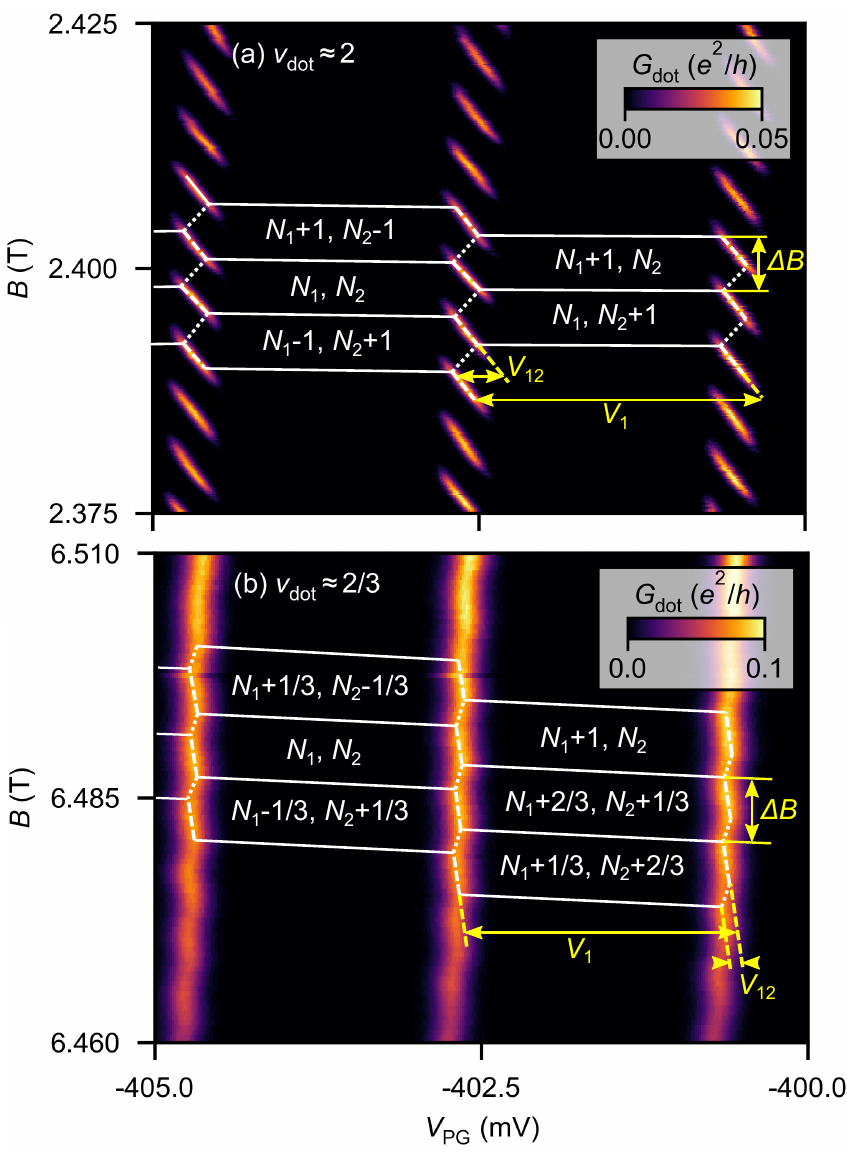}
\caption{Conductance $G_\mathrm{dot}$ as a function of the plunger gate voltage $V_\mathrm{PG}$ and the magnetic field $B$ for a dot filling factor (a) $\nu_\mathrm{dot} \approx 2$ and (b) $\nu_\mathrm{dot} \approx 2/3$. Regions of constant charge $(N_1, N_2)$ are indicated in the charge stability diagram by white lines. The charge on the outer and inner compressible region is denoted by $eN_1$ and $eN_2$, respectively. The situation for $\nu_\mathrm{dot} \approx 2/3$ in (b) allows for fractional charging of $e^* = e/3$.} 
\label{fig2}
\end{figure}

We now study the conductance of the weakly coupled \gls{QD} in the fractional quantum Hall regime for filling factor $\nu_\mathrm{dot} \approx 2/3$ and compare it to the integer quantum Hall regime at $\nu_\mathrm{dot} \approx 2$. First, we look at the conductance around filling factor $\nu_\mathrm{dot} \approx 2$ as a function of plunger gate voltage and magnetic field shown in Fig.~\ref{fig2}(a). The Coulomb resonances show a distinct periodic pattern in magnetic field that has been studied in previous works \cite{heinzel_periodic_1994, chen_transport_2009, sivan_observation_2016, liu_electrochemical_2018, roosli_observation_2020}. The pattern originates from an interplay of the two compressible regions emerging from the two filled Landau levels at filling factor 2 as schematically depicted by the light blue regions in Fig.~\ref{fig1}(a). Regions of stable charge $(N_1,N_2)$  [separated by white lines in Fig.~\ref{fig2}(a)] can be described by a capacitance model \cite{evans_coulomb_1993, rosenow_influence_2007, halperin_theory_2011, roosli_observation_2020} where $N_1$ and $N_2$ correspond to the number of electrons on the outer and inner compressible region respectively.

Changing to the fractional quantum Hall regime, the conductance depending on the magnetic field and the plunger gate voltage is shown in Fig.~\ref{fig2}(b) around filling factor $\nu_\mathrm{dot} \approx 2/3$. The Coulomb resonances of the \gls{QD} show a periodic modulation in the amplitude and the position in plunger gate voltage as a function of the magnetic field. The modulations are clearly visible while being less pronounced and extended compared to filling factor 2. The visible conductance resonances are continuously connected, in contrast to the clearly separated resonances around filling factor 2.  The observation of modulated Coulomb resonances suggests the existence of a non-trivial fractional quantum Hall state inside the \gls{QD}. Such a periodic pattern has previously not been observed for \glspl{QD} in the fractional quantum Hall regime for filling factor $\nu_\mathrm{dot} <1$ to the best of our knowledge.

To further study the modulated Coulomb oscillations at filling factor $\nu_\mathrm{dot} \approx 2/3$  and get quantitative predictions we extend the model for the integer quantum Hall effect discussed in our previous work \cite{roosli_observation_2020} to fractional filling factors. We assume the existence of two compressible regions for the fractional filling factor $\nu_\mathrm{dot} = 2/3$ separated by an incompressible $\nu_\mathrm{in} = 2/3$ region as schematically depicted in Fig.~\ref{fig1}(a) and similar to the situation at $\nu_\mathrm{dot} = 2$ where the incompressible region assumes filling factor $\nu_\mathrm{in} = 1$. In thermodynamic equilibrium the charge distribution in the \gls{QD} minimizes the electrostatic energy. Changing either the magnetic field by $\delta B$ or the plunger gate voltage by $\delta V_\mathrm{PG}$ charge imbalances arise between the two compressible regions
\begin{equation}
\begin{aligned}
\delta Q_1 &=  \Delta n_1 - \nu_\mathrm{in} \delta B \bar{A}/\phi_0 - C_1 \delta V_\mathrm{PG}/e\\
\delta Q_2 &=  \Delta n_2 + \nu_\mathrm{in} \delta B \bar{A}/\phi_0 - C_2\delta V_\mathrm{PG}/e.
\end{aligned}
\label{eq:imbalance}
\end{equation}
Here $\delta Q_i$ ($i=1,2$) denotes the charge imbalance of the outer ($i=1$) and inner ($i=2$) compressible region in units of the elementary charge $e$. The $\Delta n_i$ describe discrete changes in charge of the respective region due to quasiparticle tunneling to the other compressible region or the leads. In the fractional quantum Hall regime, this can take fractional values corresponding to a multiple of the fractional charge $e^*$ for tunneling events between the compressible regions and is not required to be an integer number of the elementary charge $e$. Upon changing the magnetic flux through the area $\bar{A}$ enclosed by the incompressible stripe at $\nu_\mathrm{in} = 2/3$, a Hall current $ \nu_\mathrm{in} \delta B \bar{A}/\phi_0$ will flow from the outer to the inner compressible region. Consequently, the (fractional) charge $\nu_\mathrm{in}e$ will be shifted when adding one flux quantum $\phi_0 = h/e$, not necessarily corresponding to the quasi-particle charge $e^*$. The plunger gate couples to the compressible regions over the effective capacitances $C_i > 0$. The change in total electrostatic energy can then be calculated to be
\begin{equation}
\delta E = \frac{1}{2}K_1\delta Q_1^2 + \frac{1}{2}K_2\delta Q_2^2 + K_{12}\delta Q_1\delta Q_2,
\label{eq:deltaE}
\end{equation}
where the $K_i\,(i=1,2)$ describe the charging energies of the compressible regions and $K_\mathrm{12}$ the cross-charging energy due to capacitive coupling between the compressible regions. This model predicts hexagonally shaped regions of stable charge as a function of the magnetic field and plunger gate voltage as the system minimizes the energy functional [Eq.~\eqref{eq:deltaE}] by assuming suitable $\Delta n_i$.

\begin{figure}[tbp]
\includegraphics[width=\columnwidth]{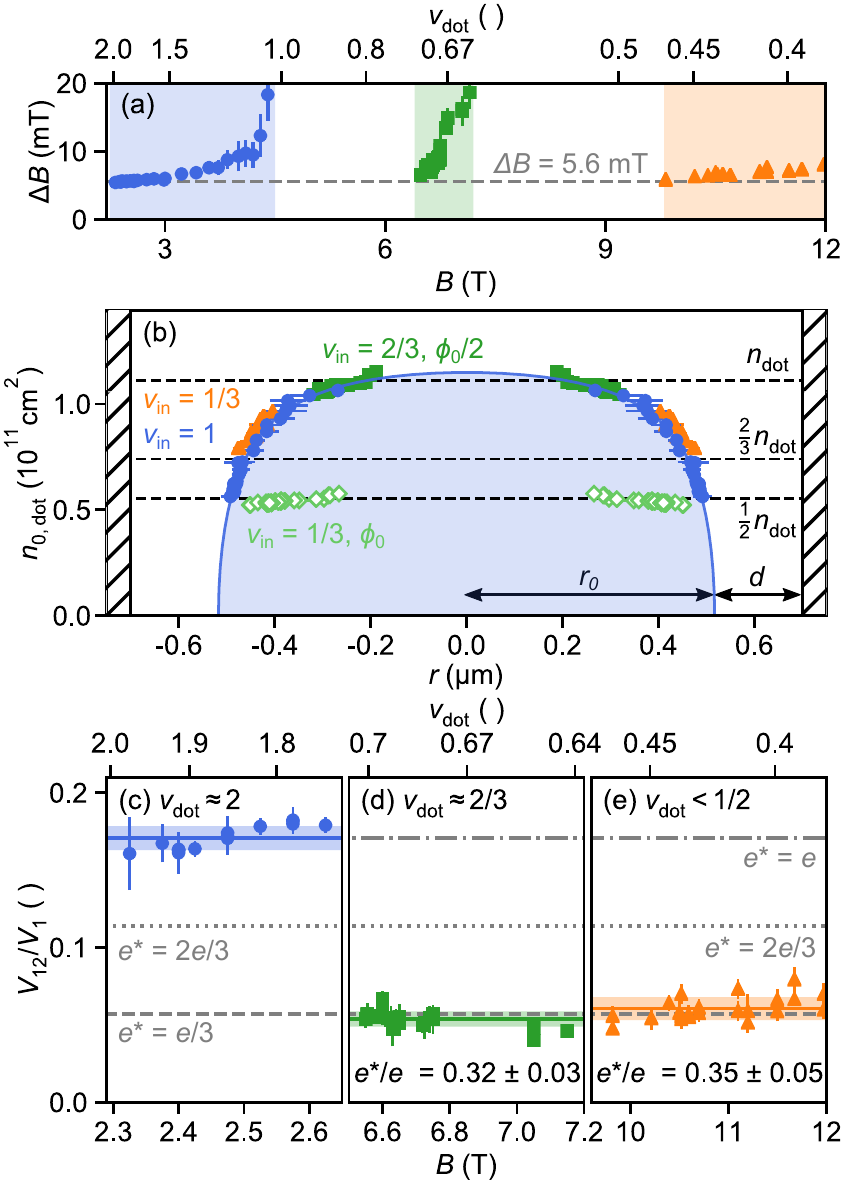}
\caption{(a) Magnetic field period $\Delta B$ indicated in Fig.~\ref{fig2}(a) and (b) as a function of magnetic field $B$. A periodic modulation is only observed for the shaded magnetic field regions. (b) Zero-magnetic-field density $n_\mathrm{0, dot}$ of the \gls{QD} as a function of the radius $r$ calculated from the magnetic field periodicities in (a) according to Eq.~\eqref{eq:radius} and \eqref{eq:density}, mirrored around $r=0$. We assume a  magnetic field periodicity $\Delta B$ corresponding to a flux quantum $\phi_0$ for $2 > \nu_\mathrm{dot} > 1$ [(c), blue dots] and  $\nu_\mathrm{dot} \gtrsim 1/3 $ [(e), orange triangles], and $\phi_0/2$ for  $\nu_\mathrm{dot} \approx 2/3$ [(d), green squares]. Assuming  a flux quantum periodicity $\phi_0$ and an incompressible stripe at $\nu_\mathrm{in} = 1/3$ for $\nu_\mathrm{dot} \approx 2/3$ instead (light green empty diamonds) the calculated charge distribution does not agree with the data of other filling factors. The blue line shows a fit according to Eq.~\eqref{eq:density_fit}. (c)-(e) Ratio $V_{12}/V_1$ of the plunger gate voltages indicated in Fig.~\ref{fig2}(a) and (b) depending on $B$ for dot filling factors around (c)  $\nu_\mathrm{dot} \approx 2$, (d) $\nu_\mathrm{dot} \approx 2/3$ and (e) $1/3 < \nu_\mathrm{dot} < 1/2$. Quasi-particle charge ratios $e/e^*$ calculated by Eq.~\eqref{eq:fractional_charge_ratio} are indicated.  } 
\label{fig3}
\end{figure}

\begin{figure*}[tbp]
\includegraphics{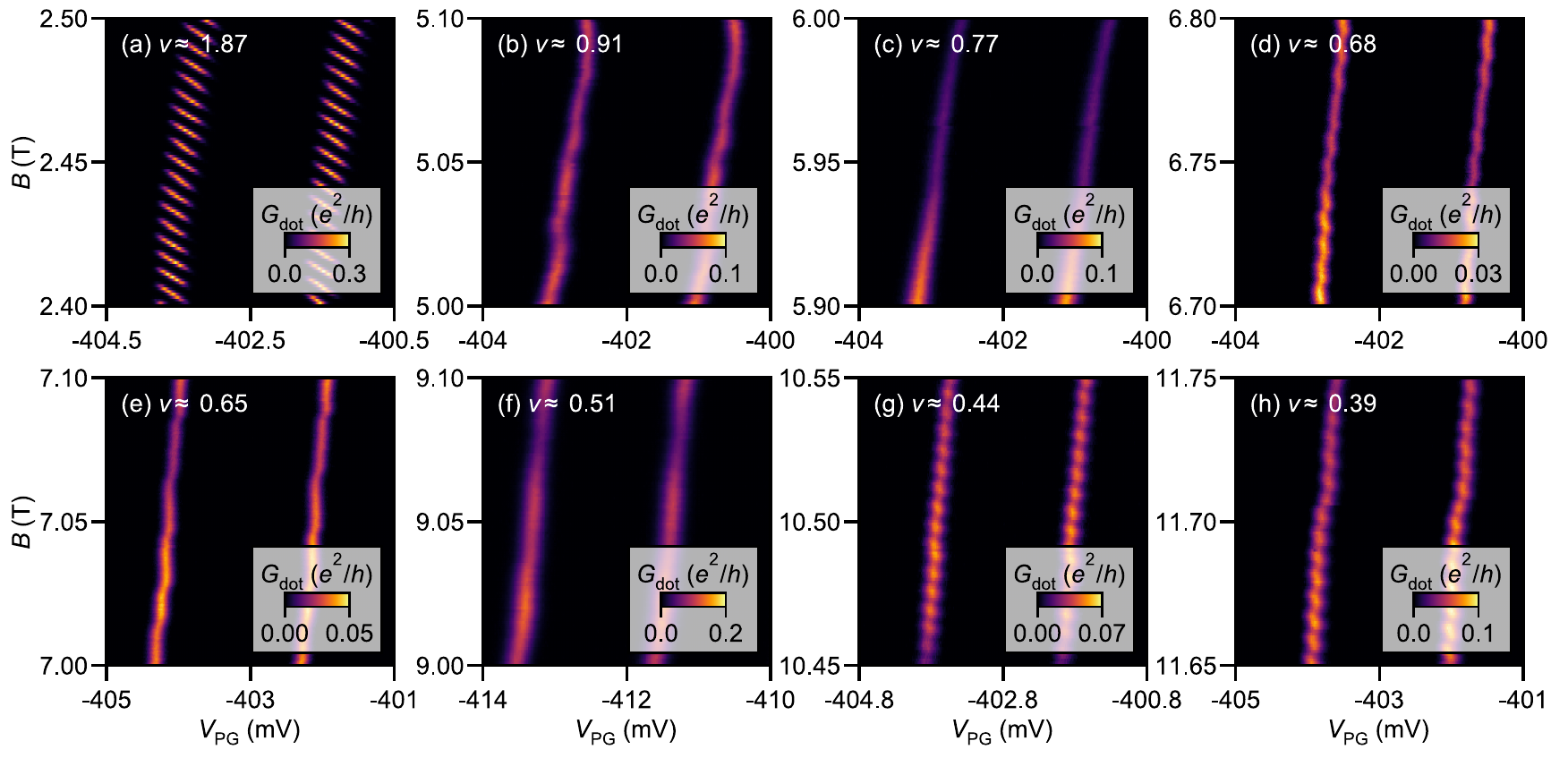}
\caption{Conductance $G_\mathrm{dot}$ as a function of the plunger gate voltage $V_\mathrm{PG}$ and the magnetic field $B$ for decreasing dot filling factors $2 > \nu_\mathrm{dot} > 1/3$. Periodic modulation of the Coulomb peaks are observed in (a), (d), (e), (g) and (h) corresponding to filling factors in the regions marked in  Fig.~\ref{fig3}(a). No periodic modulations are observed in (b), (c) and (f). } 
\label{fig4}
\end{figure*}

We interpret the data in Fig.~\ref{fig2}(b) at $\nu_\mathrm{dot} \approx 2/3$ according to this model and draw the charge stability diagram. We will now look at charging events where a fractional charge $e^*$ is rearranged between the two compressible regions, i.e., $\Delta n_1 = +e^*/e, \Delta n_2 = -e^*/e $. The corresponding magnetic field spacing $\Delta B$ coincides with the magnetic field period as indicated in Fig.~\ref{fig2}(b). The measured magnetic field period $\Delta B$ is shown in Fig.~\ref{fig3}(a) as a function of magnetic field $B$ for the regions where periodic modulations are observed. We find a stable period of $\Delta B = \SI{5.6}{mT}$ at $\nu_\mathrm{dot} \approx 2$ (blue dots) slowly rising towards and diverging at $\nu_\mathrm{dot} \approx 1$ which is directly related to the decreasing area enclosed by the incompressible $\nu_\mathrm{in} = 1$ stripe as the upper spin-split branch of the lowest Landau level is depopulated \cite{roosli_observation_2020}. For $\nu_\mathrm{dot} \approx 2/3$ (green squares) the period quickly increases with increasing $B$ as well. Interestingly, periodic modulations of the conductance are also observed for filling factors $1/2 >\nu_\mathrm{dot} >1/3$ [see Fig.~\ref{fig4}(g), (h)]  similar to $\nu_\mathrm{dot} \approx 2/3$ in Fig.~\ref{fig2}(b). In this region, the quantum dot exhibits two compressible regions separated by an incompressible region at $\nu_\mathrm{in} = 1/3$. The corresponding magnetic period for $1/2 >\nu_\mathrm{dot} >1/3$ (orange triangles) shows a slow increase similar to the behavior close to $\nu_\mathrm{dot} \approx 2$.

The magnetic field spacing between two rearrangements resulting from the model is $\Delta B = (e^*/e)\phi_0/(\nu_\mathrm{in}\bar{A})$. Assuming a circular \gls{QD} we can calculate the radius (center) of the incompressible region according to
\begin{equation}
r(B) =  \sqrt{\frac{e^*/e}{\nu_\mathrm{in}}\frac{\phi_0}{\Delta B \pi}}
\label{eq:radius}
\end{equation}
with a density in the incompressible region
\begin{equation}
n(r,B) =  \frac{eB}{h}\nu_\mathrm{in}.
\label{eq:density}
\end{equation}
Using these two equations enables us to reconstruct the zero-magnetic-field charge density distribution $n_\mathrm{0, dot}$ in the \gls{QD} from the measured magnetic period $\Delta B(B)$ in Fig.~\ref{fig3}(a). We assume fractional charge $e^* =e/3$ for $\nu_\mathrm{dot} \gtrsim 1/3$ and $2/3$ and charge $e$ for $\nu_\mathrm{dot} \approx 2$ while having an incompressible region at $\nu_\mathrm{in} = 1/3$, $2/3$ and $1$, respectively. This results in the dot density $n_\mathrm{0, dot}$ shown in Fig.~\ref{fig3}(b) with a flux quantum periodicity $\phi_0$ for $\nu_\mathrm{dot} \approx 2$ (blue dots) and $\nu_\mathrm{dot} \gtrsim 1/3$ (orange triangles), and a half-flux quantum periodicity $\phi_0/2$ for $\nu_\mathrm{dot} \approx 2/3$ (green squares). All three different regimes line up to form a smooth radial density dependence. The slightly higher radii for  $\nu_\mathrm{dot} \gtrsim  1/3$ probably reflect the lower voltages applied to the barrier gates in this regime. For $\nu_\mathrm{dot} \approx 2/3$ we can exclude periodic modulations spaced by a full flux quantum $\phi_0$ that would originate from an incompressible region of $\nu_\mathrm{in} = 1/3$. They would lead to the light green, empty diamonds in Fig.~\ref{fig3}(b) which do not line up with the points of the other regimes.

In a study of an anti-dot embedded into a $\nu=2/3$ FQH state  \cite{kou_coulomb_2012}, a flux period of $\phi_0$ was found. This experiment was analyzed with the help of an electrostatic model which assumed that the edge of the $\nu=2/3$ quantum Hall state consists of a downstream propagating integer channel and an upstream propagating fractional $1/3$-channel \cite{levy_schreier_thermodynamic_2016}. Depending on the strength of the Coulomb repulsion between these channels, the flux periodicity was found to be $\phi_0$ for weak coupling, and 
$\phi_0/2$ for strong coupling. In the latter case, the two channels can be described as a single compressible region, as in our model 
Eq.~\eqref{eq:deltaE}. In the present experiment, we cannot distinguish any additional $\phi_0$ period.

We can fit the dot density $n_\mathrm{0, dot}(r)$ in Fig.~\ref{fig3}(b) using a model proposed by Lier and Gerhardts \cite{lier_self-consistent_1994} for the position of incompressible stripes at the gated  edge of quantum Hall systems. The presence of a charged gate leads to a reduction of the bulk density $n_\mathrm{bulk}$ by a factor $s$. To calculate the density distribution we assume two gates placed symmetrically around the center at $r=0$ resulting in
\begin{equation}
\begin{aligned}
n_\mathrm{0, dot}(r) &=  n_\mathrm{bulk} s(r,r_0,d)s(r,-r_0,-d) \quad \mathrm{with}\\
s(r,r_0,d) &= \sqrt{(r_0-r)/(r_0+d-r)},
\end{aligned}
\label{eq:density_fit}
\end{equation}
where $r_0$ is the radius where the density drops to 0 and $d$ is the depletion length around the gate. The fit (blue line) to our data at $\nu_\mathrm{dot} \approx 2$ is shown in Fig.~\ref{fig3}(b) with $r_0 = \SI{517 \pm 1}{nm}$ and $d = \SI{131 \pm 4}{nm}$. This agrees well with the lithographic dot size $r_\mathrm{lith} \approx \SI{0.7}{\micro m} \approx r_0 + d$.

From electrostatic simulations using \textit{COMSOL} we calculate a magnetic field period $\Delta B = \SI{6.3}{mT}$ at a magnetic field of $ B = \SI{10}{T}$ for filling factor $\nu_\mathrm{dot} \gtrsim  1/3$ and gate voltages comparable to the experimentally applied values, which is in good agreement with the experimentally determined value. Similary we get $\Delta B = \SI{11.6}{mT}$ at a magnetic field of $ B = \SI{7}{T}$ for filling factor $\nu_\mathrm{dot} \approx  2/3$, again in good agreement with the experiment. The calculated total charge on the \gls{QD} corresponds roughly to $N_\mathrm{tot}^{(\mathrm{sim})} = 870$ electrons. This is comparable to the experimentally derived value of $N_\mathrm{tot}^{(\mathrm{exp})} = 820$  from experimental values for the density $n_\mathrm{dot}$ and the area $\bar{A}$. The results of the electrostatic simulations agree well with the experimental observations.

Returning to the charge stability diagram in Fig.~\ref{fig2}(b), we calculate within our model the slope of an internal charging line between the two compressible regions
\begin{equation}
\left.\frac{\delta B}{\delta V_\mathrm{PG}}\right|_\mathrm{rearr.} = -\dfrac{1}{\nu_\mathrm{in}} \frac{\phi_0 e (\alpha_1 - \alpha_2)}{\bar{A}[(K_1 - K_{12}) + (K_2 - K_{12})]} < 0,
\label{eq:slope_rearrangement}
\end{equation}
where $\alpha_1 = (K_1C_1 + K_{12}C_{2})/e^2$ and $\alpha_2 = (K_{12}C_1 + K_{2}C_{2})/e^2$ denote the lever arms of the plunger gate on the respective compressible region. This slope is negative as generally $\alpha_1 > \alpha_2$ which limits the ways we can draw the recharging lines connecting the visible Coulomb oscillations. Additionally, we calculate the slopes corresponding to a constant charge on either of the two compressible regions 
\begin{equation}
\begin{aligned}
\left. \frac{\delta B}{\delta V_\mathrm{PG}}\right|_{\Delta n_1 = 0} &= - \dfrac{1}{\nu_\mathrm{in}}C_1\phi_0/(e\bar{A}) &< 0,\\
\left. \frac{\delta B}{\delta V_\mathrm{PG}}\right|_{\Delta n_2 = 0} &= \dfrac{1}{\nu_\mathrm{in}}C_2\phi_0/(e\bar{A}) &> 0.
\end{aligned}
\label{eq:slope_n1n2const}
\end{equation}
These constraints define the orientation of hexagons in the charge stability diagram as shown in Fig.~\ref{fig2}(b). Crossing any segment of the hexagon boundary with non-zero conductance  in $V_{\mathrm{PG}}$-direction in Fig.~\ref{fig2}(b), the total charge on the \gls{QD} changes by $e$. We see that crossing the boundary segment with negative slope implies $\Delta n_1 = 1$ and $\Delta n_2=0$, while crossing it along a segment with positive slope corresponds to $\Delta n_1 = 2/3$ and $\Delta n_2 = 1/3$. The charge of an electron tunneling into the \gls{QD} can therefore be split among the compressible regions into fractional quasiparticle charges.  This could be the reason why the Coulomb resonances are connected for $\nu_\mathrm{dot} \approx 2/3$ [Fig.~\ref{fig2}(b)]. At $\nu_\mathrm{dot} \approx 2$  on the other hand, the Corbino conductance across the incompressible region might be very small, such that electron tunneling into the inner compressible region is not significant on transport time scales, giving rise to clearly separated resonances  [Fig.~\ref{fig2}(a)].

Analyzing the charge stability diagram for the integer and fractional quantum Hall regimes in Fig.~\ref{fig2} further, we extract an experimental value for the quasi-particle charge for fractional filling factors $\nu_\mathrm{dot} = 2/3$ and $1/3$. To this end we calculate from the model the voltage differences $V_1 = K_1/(e\alpha_1)$ and $V_{12} = (e^*/e)(K_1-K_{12})/(e\alpha_1)$ which appear in the measurements in Fig.~\ref{fig2}(a)~and~(b) as separations of visible charging lines as indicated. By taking the ratio $V_{12}/V_1$ we eliminate the lever arm $\alpha_1$. Assuming that the ratios $K_1/K_{12}$ are independent of the filling factor, the quasiparticle charge is then calculated by comparing the ratios for the integer and fractional filling factors
\begin{equation}
\frac{e^*}{e} = \frac{V_{12}^{(\mathrm{fract})}/V_1^{(\mathrm{fract})}}{V_{12}^{(\mathrm{int})}/V_1^{(\mathrm{int})}}.
\label{eq:fractional_charge_ratio}
\end{equation}
Figures~\ref{fig3}(c)-(e) show the ratio $V_{12}/V_1$ as a function of the magnetic field for filling factors around $\nu_\mathrm{dot} \approx 2$, $\approx 2/3$ and $\gtrsim 1/3$. Each point corresponds to a measurement  as depicted in Figs.~\ref{fig2}(a)~or~(b) centered around a certain magnetic field value. The extracted relevant parameters are averaged over several hexagons of such a charge stability diagram. The ratio $V_{12}/V_1$ is roughly constant within each filling factor regime. We extract a fractional charge $e^*/e = \SI{0.32 \pm 0.03}{}$ and $e^*/e = \SI{0.35 \pm 0.05}{}$, respectively, by comparison to $\nu_\mathrm{dot} \approx 2$. This indicates a fractional charge $e^* = e/3$ for quasiparticles tunneling in the \gls{QD} for both $\nu_\mathrm{dot} \approx 2/3$ and $\gtrsim 1/3$. For $\nu = 1/3$, fractional charge $e^* = e/3$ has previously been found from shot noise measurements \cite{de-picciotto_direct_1997,saminadayar_observation_1997}, localized states \cite{martin_localization_2004} or quantum Hall Fabry-P\'{e}rot interferometers \cite{nakamura_aharonovbohm_2019}, while at $\nu= 2/3$ different experiments indicated different values, namely $e/3$ \cite{bid_shot_2009}, $2e/3$ \cite{bid_shot_2009, kou_coulomb_2012} and $e$ \cite{nakamura_aharonovbohm_2019}.

In order to give an overview of the evolution of the conductance measurements for decreasing filling factors $2 > \nu_\mathrm{dot} > 1/3$, we show additional data in Fig.~\ref{fig4}. As indicated in Fig.~\ref{fig3}(a), periodic modulations of the Coulomb resonances are only observed within the marked regions. The evolution from the distinct pattern at $\nu_\mathrm{dot} = 2$ in Fig.~\ref{fig4}(a) towards filling factor 1 has been studied in detail in previous work \cite{roosli_observation_2020} and is found here to behave in the same way. For $1> \nu_\mathrm{dot} > 2/3$ the Coulomb resonances show no periodic modulations as seen in Figs.~\ref{fig4}(b)~and~(c). In the fractional quantum Hall regime the Coulomb resonances are modulated around filling factor $\nu_\mathrm{dot} \approx 2/3$ and $1/2>\nu_\mathrm{dot}$ as shown in Figs.~\ref{fig4}(d),~(e),~(g)~and~(h). The two regions are interrupted by a region around filling factor $\nu_\mathrm{dot}=1/2$ where no modulations are observed [see Fig.~\ref{fig4}(f)].

In conclusion, we have studied the magneto-conductance of a \SI{1.4}{\micro m}-wide \gls{QD} in the fractional quantum Hall regime for filling factors $\nu < 1$. Around $\nu_\mathrm{dot} \approx 2/3$ and $1/2>\nu_\mathrm{dot}>1/3$ we observe periodic modulations of Coulomb resonances as a function of magnetic field. Assuming two compressible regions separated by an incompressible stripe at $\nu_\mathrm{in} = 2/3$ and $\nu_\mathrm{in} = 1/3$, respectively, we have successfully used an electrostatic model to describe the phase diagram as a function of magnetic field and plunger gate voltage. We extract the charge density distribution of the \gls{QD} at zero magnetic field. By comparing our measurements in the fractional regime with measurements at $\nu_\mathrm{dot} \approx 2$, we find fractional Coulomb blockade between the compressible regions in the \gls{QD} with quasiparticle tunneling of fractional charge $e^*/e = \SI{0.32 \pm 0.03}{}$ and $e^*/e = \SI{0.35 \pm 0.05}{}$ for the two fractional regimes respectively. Quantum dots and quantum Hall Fabry-P\'{e}rot interferometers have been shown to be closely related in the integer quantum Hall regime \cite{rosenow_influence_2007,stern_interference_2010,halperin_theory_2011, sivan_observation_2016, roosli_observation_2020}. We have demonstrated experimentally that this relation persists in the fractional regime. While interferometry experiments in the fractional quantum Hall regime have been shown to be very intricate and sensitive \cite{ofek_role_2010, sabo_edge_2017, nakamura_aharonovbohm_2019, willett_interference_2019}, fully Coulomb blockaded devices may provide an alternative experimental approach. Our observations complement and extend interferometry experiments of fractional quantum Hall states.

\begin{acknowledgments}
The authors acknowledge the support of the ETH FIRST laboratory and the financial support of the Swiss Science Foundation (Schweizerischer Nationalfonds, NCCR QSIT). B.~R. would like to acknowledge support by DFG grant RO 2247/11-1.
\end{acknowledgments}

%

\end{document}